# Neutron lifetime, dark matter and search for sterile neutrino

A. P. Serebrov[*], R.M. Samoilov , I.A. Mitropolsky , A.M. Gagarsky

*Petersburg Institute of Nuclear Physics , NRC"Kurchatov institute",*

*188300 Gatchina, Leningrad region, Russia*

Abstract

A review is focused on experimental measurements on neutron lifetime. The latest measurements with a gravitational trap (PNPI NRC KI) and a magnetic trap (LANL, USA) confirmed PNPI result of 2005. The results of measurements with storage of ultra cold neutrons are in agreement, yet, there is discrepancy with a beam experiment by 3.5σ (1% of decay probability), which is discussed in literature as "neutron anomaly" along with the ideas of explaining it by decay into dark matter partially.

The second part of the paper is devoted to so called "reactor antineutrino anomaly", which refers to deficiency of the measured flux of antineutrino from reactor in respect to the calculated flux by 3σ (deviation by 6.6%). The issue is under extensive debate at neutrino conferences, some experiments being conducted on search for a sterile neutrino, i.e. transition into dark matter in neutrino sector.

Specific feature of the proposal in this paper lies in the fact that both anomalies can be accounted for by one and the same phenomenon of oscillation in baryon sector between a neutron and a neutron of dark matter $n \to n'$ with mass $m_{n'}$, somewhat less than mass $m_n$ of an ordinary neutron. Calculations of the proposed model require one free parameter: mass difference $m_n - m_{n'}$. If one normalizes probability of $n \to n'$ oscillations for a free neutron on "neutron anomaly" 1%, then, having succeeded to interpret 6.6% of neutron anomaly in calculations, one can determine mass difference and thus, neutron mass of dark matter neutron. According to preliminary estimations, the mass difference is $m_n - m_{n'} \approx$ 3MeV. However, the analysis of cumulative yields of isotopes occurs in fission fragments was performed and it does not confirm possibility of existence of additional decay channel with emission of dark matter neutron with mass difference $m_n - m_{n'} \approx$ 3MeV.

The result of the analysis is the conclusion that for mirror neutrons the region of the mass difference $m_n - m_{n'} \geq 3$ MeV is closed. The region of the mass difference $m_n - m_{n'} \leq 2$ MeV turned out to be not closed, because there are practically no nuclides with neutron binding energies below 2 MeV.

[*]A.P.Serebrov, E-mail: serebrov_ap@pnpi.nrcki.ru

## 1. History of neutron lifetime measurements.

Neutron lifetime is one of the most important fundamental constants for weak interaction theory and cosmology. Neutron has the longest lifetime among the unstable elementary particles, its lifetime is ~880 seconds. It is the great length of lifetime, i.e. very small decay probability, to be the reason that parameter is very hard to measure. For example, in a cold neutron beam at 1 m distance only one of a million neutrons passing through experimental setup occurs to decay. However, there is an alternative way to measure neutron lifetime using ultracold neutrons (UCN). These neutrons have very low kinetic energy, they are reflected from the walls of material traps and magnetic traps with magnetic field gradients at the wall. The idea of the experiment is to store neutrons into the trap and observe their decay. The loss probability in the trap can be decreased to 1-2% of the neutron decay probability, applying the cryogenic material traps [1,2] and even lower losses are achievable with magnetic traps [3,4,5]. That means neutrons can be stored in traps and neutron lifetime can be measured almost directly, introducing small corrections for UCN losses in the trap.

The history of neutron lifetime measurements covers the significant period of time starting from the first experiments in 70s at neutron beams [6,7]. Since then the accuracy of measurements has increased over an order of magnitude, and significant progress was achieved using UCN.

However, the progress in UCN method was not as certain as it might seems. The first experiments with UCN storing lacked accuracy due to small UCN density into a trap [8]. Accuracy of the experiments increased after the UCN sources with high intensity were created in Gatchina [9] and Grenoble [10]. The significant success was achieved by using fluorine containing oil (fomblin), where the Hydrogen atoms replaced by fluorine [11, 12]. However, the probability of UCN losses in those experiments was ~30% [11] and ~13% [12] of neutron decay probability. The experimental problem was the extrapolation of UCN trap storage time to neutron lifetime, performing measurements with various collision frequencies using various trap geometry. Extrapolation range was about 200s [11] and 100s [12], hence achieving 1 s accuracy for extrapolation was an extremely difficult task.

Besides, the effect of low energy heating was discovered and it leads to systematic effect in neutron lifetime measurements [13-16]. UCN measurements of neutron lifetime were significantly improved by applying an open-topped cryogenic trap where neutrons are trapped by Earth gravity [17]. Using low temperatures the effects of inelastic scattering and small heating were suppressed and loss probability at walls became about 1÷2% of neutron decay probability. Here the extrapolation range becomes only 5 ÷ 10s. That way the accuracy of 1s for neutron lifetime is achievable.

Within the experiment carried out in 2004 in ILL by the collaboration of PNPI and JINR[1] was obtained the neutron lifetime 878.5±0.7±0.3s, here the first error is statistical and the second is systematical. The result of experiment carried out in Gatchina [16] with similar trap was in good agreement within accuracy and the difference was less than 2σ. Neutron lifetime value in PDG 2006 was 885.7±0.8s. The discrepancy between the result of the new experiment carried out in 2005[1] and PDG value was 6.5σ and caused a wide discussion with significant mistrust in that discrepancy. However, in two years in the first experiments with magnetic trap with permanent magnets [3,4] the result was confirmed and the measured lifetime was 878.2±1.9s.

In 2010 in the experiment MAMBO II [18] was obtained the result 880.7±1.8 c. In 2012 the results of the experiments with room temperature fomblin [11,12] were corrected to be 882.5±2.1s[19] and 881.6±2.1s[20]. In 2015 the new experiment was carried out by scientific group led by V.I.Morozov and the result was 880.2±1.2s [21]. Our scientific group (PNPI, Gatchina) have developed a project of the experiment with big gravitational trap in 2010 to check the result of our experiment carried out in 2005 [22]. This experiment with big gravitational trap was carried out by PNPI NRC KI-ILL-RAL collaboration in 2017[2]. The obtained result is 881.5±0.7±0.6 so both results are in agreement within 2σ. Also in 2017 was published the result of experiment LANL [5] with magnetic trap of UCN and obtained result is 877.7±0.7±0.3s. Summarizing all those results we can conclude that the result obtained in 2005 is confirmed by the experiments with UCN storing. The historical diagram of measurements is shown in figure 1.

In figure 2 the diagram of neutron lifetime measurements is presented starting from 2005. On the left one can see the results of the experiments with UCN storing in material and magnetic traps. From the data one can conclude that the results of storing experiments are consistent within two standard deviations. On the right are the results from neutron beams with proton trap, which significantly differs [23, 24]. In the table are listed the results of the experiments with statistical and systematic errors and also the total error calculated as a linear sum of errors. Notice that we use linear sum of statistical and systematic errors, which is more conservative way than square addition.

The discrepancy between beam [23],[24] and UCN storing experiments is 3.5σ if we use quadratic addition and 2.6σ is we use linear addition. In any case it is a noticeable discrepancy [25] and it is sometimes called "neutron anomaly"[26]. It would be very interesting to have the results of repeated experiment with neutron beam and proton trap, and also an independent experiment with neutron beam and registration of both protons and electrons from neutron decay. The repeating of the experiment with proton trap is planned as well as a new experiment at neutron beam[27]. It may clarify the neutron anomaly problem or will lead to more certain proofs of its existence.

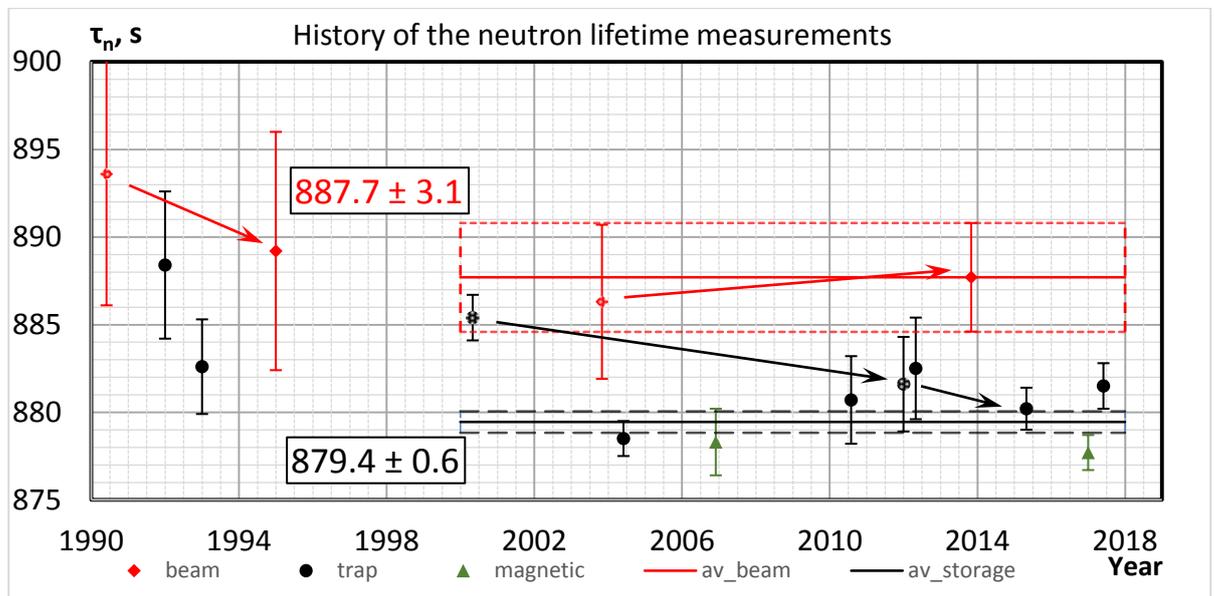

Fig. 1.Experimental results of neutron lifetime measurements from 1990, the discrepancy of the results obtained in 2005 [1]and 2000 [11], the correction of liquid fomblin experiments [18,19] and the new experiment [20], finally, the new results of 2017 [22] and [2].

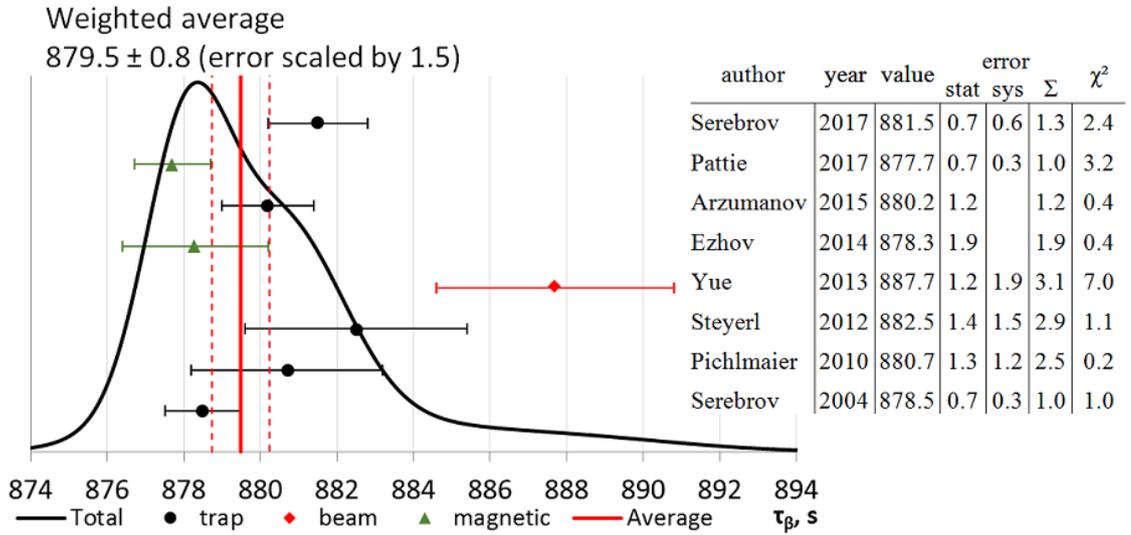

Fig. 2. Neutron lifetime measurements diagram from 2005 in experiments with storing UCN in magnetic and material traps, and also in the neutron beam experiment with proton trap to detect neutron decay protons.

## 2. Measurements of neutron decay asymmetry and Standard Model test.

Consider in details the researches of neutron decay including measurements of asymmetry of β-decay and tests of SM. It is well-known that CKM matrix element $V_{ud}$ can be determined from β-decay by measuring neutron lifetime and decay asymmetry (Fig. 3) and the result can be compared with other methods of $V_{ud}$ calculations. In fig. 4 are presented the results of tests of neutron β-decay data for determining element $V_{ud}$ using the ratio of axial and vector weak interaction coupling constants ($G_A/G_V=\lambda$), taken from PDG 2017 and the most precise measurements of electron asymmetry of β-decay[28].

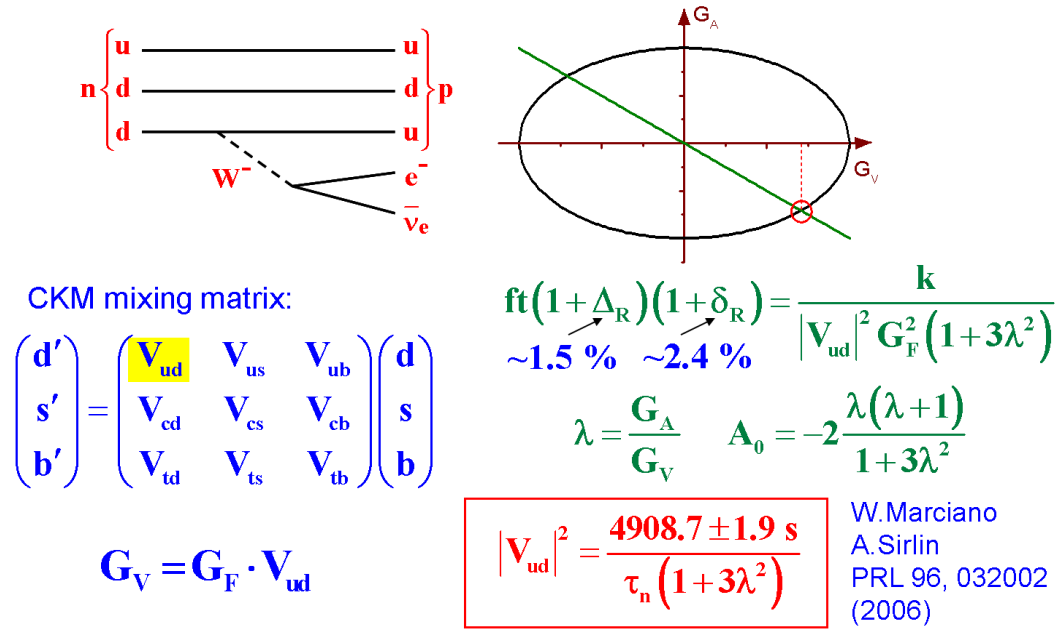

Fig. 3. Calculation of matrix element $V_{ud}$.

The value of $V_{ud}$ from β-decay is obtained by observing the intersection of data from $\tau_n$ and $\lambda=G_A/G_V$, it can be compared with $V_{ud}$ value obtained from superallowed $0^+ \to 0^+$ nuclear transitions and with $V_{ud}$ value obtained from CKM matrix unitarity ($V_{ud}^2 + V_{us}^2 + V_{ub}^2 = 1$).

One can infer that SM test is successful only with usage of neutron lifetime value from UCN storing experiments and joint usage of the most precise data of β-decay asymmetry [28]. In this situation, we have to conclude that it is necessary to carry out new experiments for measuring $\lambda=G_A/G_V$ from β-decay to confirm the most precise result [28]. On the other hand, in neutron decay research there is still space to make hypothesis beyond SM.

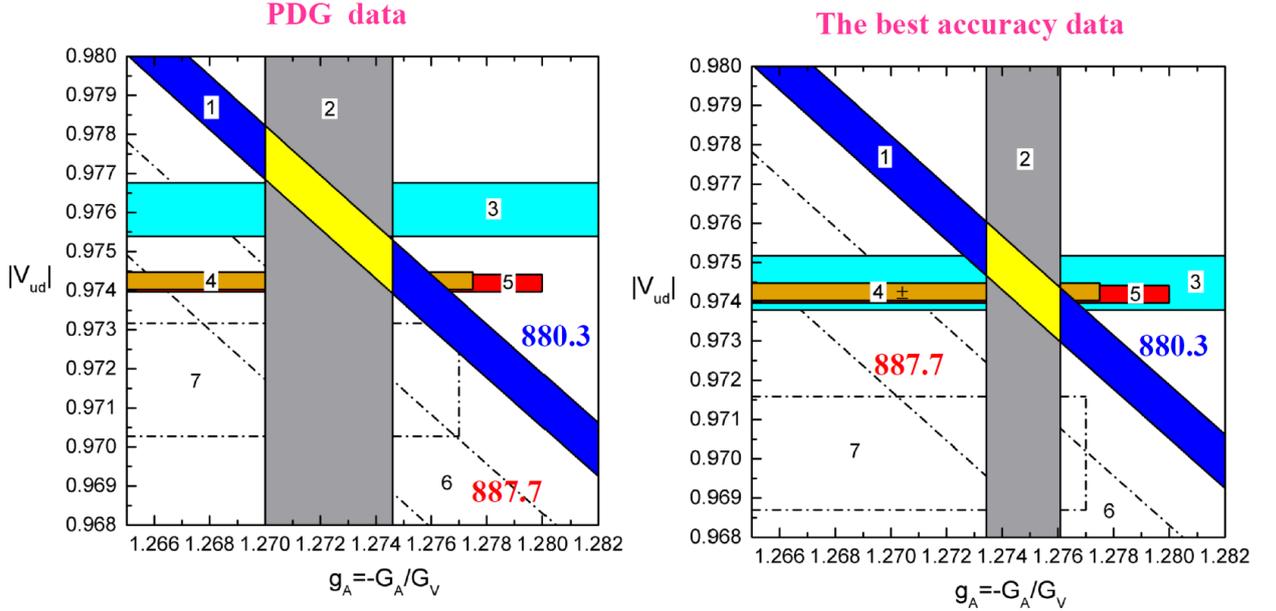

Dependence of the CKM matrix element $|V_{ud}|$ on the values of the neutron lifetime and the axial coupling constant $g_A$. (1) neutron lifetime, PDG 2015 (w/o Yue 2013); (2) neutron β-asymmetry, PDG 2015; (3) neutron β-decay, PDG 2015; (4) unitarity; (5) $0^+\rightarrow 0^+$ nuclear transitions; (6) neutron lifetime, Yue 2013; (7) neutron β-decay, Yue 2013 + PDG 2015.

Dependence of the CKM matrix element $|V_{ud}|$ on the values of the neutron lifetime and the axial coupling constant $g_A$. (1) neutron lifetime, PDG 2015 (w/o Yue 2013); (2) neutron β-asymmetry, PERKEO II; (3) neutron β-decay, PDG 2015 (w/o Yue 2013) + PERKEO II; (4) unitarity; (5) $0^+\rightarrow 0^+$ nuclear transitions; (6) neutron lifetime, Yue 2013; (7) neutron β-decay, Yue 2013 + PERKEO II.

Fig. 4. The analysis of neutron lifetime value from beam experiment 887.7 ±2.2 and UCN storing experiments 880.3±1.0 with asymmetry data from PDG.

### 3. n→n′ oscillation hypothesis

It is conventional to start a serious discussion about an effect or experimental data discrepancy than it reaches 5σ, e.g. resonances in high energy physics. However, usually the discussion on ideas starts from 3σ level considering it enough to start an analysis of the interpretation, yet it is not source of final conclusions.

When in 2005 was obtained the result 878.5±0.7±0.3s [1] having a 6.5σ deviation from PDG the one of ideas to explain it was the oscillations n→n′(neutron - mirror neutron[29]).

Our world is "left" corresponding to weak interaction and the idea of global symmetry restoration is under discussion for long time [30]. Aiming on restoring global symmetry one can assume that dark matter world is "right" corresponding to space inversion. In the simple scheme with "mirror Standard Model" the mirror neutron n' is the dark matter particle with the same mass as neutron but with the opposite value of magnetic moment and very small constant of interaction with usual matter, yet the gravitational interactions are the same. In that case n→n′ transitions are possible in absence of magnetic fields (both usual magnetic fields and mirror magnetic fields of dark matter). After the transition the mirror neutron leaves the trap area for it has almost zero interaction with matter. In that case neutron lifetime measured in the UCN storing experiments would become smaller. The idea of possible n→n′ oscillation was put forward in work [29] in 2006. Experimental searches for n→n′ oscillation were made in works [31-34]. The best restriction on the oscillation was obtained in [32]. It was shown that oscillation period exceed 414s (90% C.L.) or the oscillation probability is less than $2.4\cdot 10^{-3}$ $s^{-1}$ in the absence of magnetic field. In 2009 the upper bound on the n→n′ oscillation period was increased to 448 s (90% C.L.)[33]. The n→n′ oscillations were not observed, yet the hypothesis of mirror transitions were not closed.

It seems impossible to observe or totally exclude transitions n→n′ in the presence of mirror magnetic field with unknown value and direction, especially if it time dependent. Furthermore, if mirror neutron mass significantly differs from neutron mass than to compensate this energy difference for n→n′ transition it requires nuclear fields, not magnetic. That means the transitions n→n′ can occur in nuclei. It will be discussed later.

It is still possible that for free neutron transitions n→n′ are exists but due to mass difference the probability is much less than neutron decay probability. Total neutron decay probability can be higher than the probability of β-decay into p, e⁻ and $\bar{\nu}$. The direct assumption of the additional decay mode without proton in final state can explain the discrepancy between neutron beam and UCN storing experiments in which all decay modes should be taken into consideration. But approach to solve problems with assumptions which can not be verified experimentally are not constructive in physics. For this reason mirror neutron transition or transitions in any kind of dark matter should be put aside until there will be any ideas of experimental confirmation.

## 4. Decay into neutral dark matter

In standard neutron decay scheme there are three modes of decay, however almost all decays occurs with the proton in final state and only about 1% has γ quantum with a proton.

$n \to p + e^- + \bar{\nu}_e$    100%

$n \to p + e^- + \bar{\nu}_e + \gamma$    (9.2±0.7)·10⁻³ [35]

$n \to H + \bar{\nu}_e$    3.9·10⁻⁶ [36]

The γ quantum appears as a result of bremsstrahlung process of decay electron and its energy depends on electron energy by $E_\beta^{-1}$. Relative probability of this process is about 1%, but it is automatically taken into account in experiment [24] cause it has proton in final state. The process which is more suitable to concern neutron anomaly is neutron decay into hydrogen atom, which can not be hold in electro-magnetic trap in experiment[24], but it has very small relative probability of about 3.9·10⁻⁴% [36]. Yet 1% relative probability is required to explain neutron anomaly.

Recently an interesting explanation of the neutron decay anomaly was published in work [37]. It is based on introducing additional decay channel into dark matter in final state. Assuming those particles are stable in final state then they can be the dark matter particles with mass close to neutron mass. Regarding the ideas discussed above this transition into dark matter is very similar to transition into mirror neutron - dark matter particle with mass close to neutron mass. It should be noticed that in the dark matter model [37] the interaction of dark matter with baryons is assumed. In this scenario a monoenergetic photon in energy range 0.782-1.664 MeV is yielded in neutron lifetime experiment with 1% branching [37]. That is very important and reveals that experimental test is possible. That experimental test was performed. [38] almost right after the publication [37]. At 4σ confidence level monochromatic γ-quanta were not observed.

We would like to discuss the consequences of that assumption (and also the assumption of mirror neutron transition) for other processes of nuclear physics rather than discuss the details of the theoretical model.

## 5. About a possibility of connection between neutron anomaly and reactor antineutrino anomaly

Clearly, if there is a free neutron into dark matter decay channel than there should be a similar process of nuclear neutron decay. Such processes, if they exist, are ended for stable isotopes long time ago. But we can discuss unstable isotopes which occurs in fissions of $^{235}$U, $^{239}$Pu, $^{241}$Pu and $^{238}$U. Concerning that we would like to discuss the problem of "reactor antineutrino anomaly". The problem is that there is a deficiency of experimentally measured reactor antineutrino flux in comparison with calculated flux [39,40]. That deficiency is observed at 3-σ level and also can be considered only as reason for a theoretical discussion. This problem is under consideration at neutrino conferences and in literature [39-54]. New experiments for searching sterile neutrino are carried out or under preparation.

Antineutrino spectrum calculation from reactor is based on the information about β-spectrum of the isotopes occur in fission process and kinetic bound of electrons and antineutrino: $E_{\bar{\nu}} = E_0 - E_\beta$, where $E_0$ -total energy distributed between electron ($E_\beta$) and antineutrino($E_{\bar{\nu}}$). $N_{\bar{\nu}}(E_{\bar{\nu}}) = N_e(Q_\beta - E_e)$, where $Q_\beta$ is maximal energy of $\beta$-spectrum. Total antineutrino spectrum for all radionuclides in fission, e.g. $^{235}$U, is calculated by summing up all contributions from the nuclides (about a thousand) and using partial probability of β-decay. That a huge amount work which is done by big scientific groups. It assumed that all possible β-decaying nuclei are taken into account. If there are other unknown nuclei than they can only increase the effect of antineutrino deficiency. It seems like one should look for overestimation of β-electron yield in calculations.

In our analysis it is enough to consider one simple assumption that there is another partial decay for a particular radionuclide which is unobserved. Assume that the radionuclide has known probability of appearing and 100% probability of β-decay and we take its β-spectrum and calculate corresponding antineutrino spectrum. But, if we assume that there is a 1% probability of unobserved process (for example in dark matter decay channel or by any neutron decay channel uncontrolled in the experiment), than β-electron yield and corresponding antineutrino will be only 99%, which means in our calculations antineutrino flux is overestimated. As a result, relying on such direct and simple calculation, but without noticing another decay channel we obtain deficiency of experimentally observed antineutrino flux.

In that scheme antineutrino deficiency for various isotopes $^{235}$U, $^{239}$Pu, $^{241}$Pu, $^{238}$U can be different unlike that in scheme of explaining antineutrino deficiency with oscillation in sterile state [38,39]. Those schemes can be distinguished experimentally by measuring antineutrino flux dependence on range to reactor core. Deviation from $L^{-2}$ law, where L is range to reactor center would prove that there exist antineutrino oscillations, but range independent deficiency contradict the oscillation idea. For now, there are experimental data at 2.6σ level that antineutrino reactor anomaly strongly depends on concentration $^{235}$U in the reactor [41]. It is possible that reactor anomaly dose not connected with oscillation, but with processes in reactor or calculations. Confidence level is not sufficient yet for final conclusion.

If there is a missing decay channel, for instance, neutron one, than we do not need to introduce any new physics. On the other hand, the scheme with dark matter decay channel does not require the idea of neutrino oscillation.

At the moment, there are two experiments in which antineutrino flux from reactor at small distances is measured: Neutrino-4[42,43] and DANSS[44,45]. Distance range in neutrino flux from reactor measurements in Neutrino-4 is 6.5-11.5m and in DANSS experiment is 10.5 to 13m. These ranges overlap and results of the measurements can be connected in the overlapping area. That joint dependence of the antineutrino flux from the reactor is shown in fig.5.

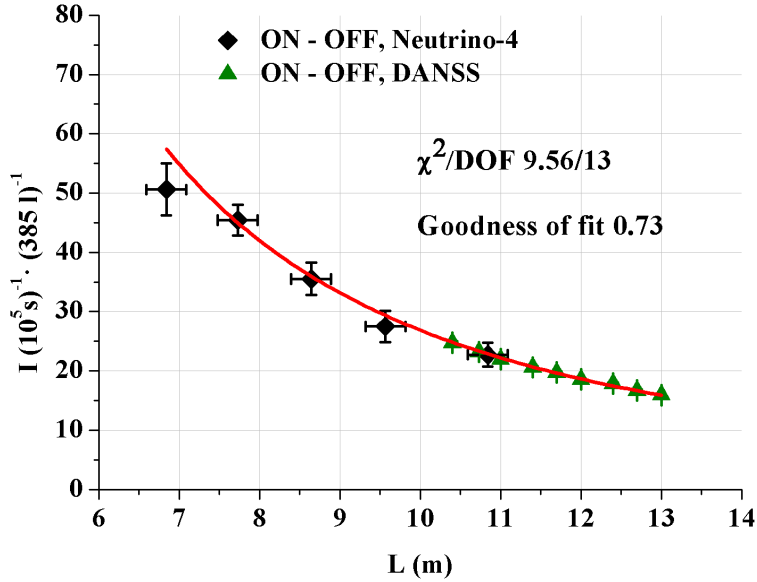

Figure 5. Reactor antineutrino flux distance dependence after combining Neutrino-4 and DANSS experimental data. Solid line is the fit for dependence $1/L^2$, where L – distance from the center of reactor core.

In figure 6 are presented the same experimental results but the $1/L^2$ decrease law is already taken into consideration to observe the deviations. Experimental results in long distance area are attached to reactor anomaly $R = 0.934 \pm 0.024$ [46], where R is the ratio of measured antineutrino flux to calculated one. Statistical errors at small range are too high to seriously discuss observing of oscillation effects. The significant increase in statistical accuracy is required in experiment Neutrino-4, and that is why the new experimental setup is developing now.

Oscillation process can be described by the following equation:

$$P(\tilde{\nu}_e \to \tilde{\nu}_e) = 1 - \sin^2 2\theta_{14} \sin^2(1.27 \frac{\Delta m^2_{14}[eV^2]L[m]}{E_{\tilde{\nu}}[MeV]}),$$

where $E_{\tilde{\nu}}$ is antineutrino energy, $\Delta m^2_{14}$ and $\sin^2 2\theta_{14}$ are the unknown oscillation parameters.

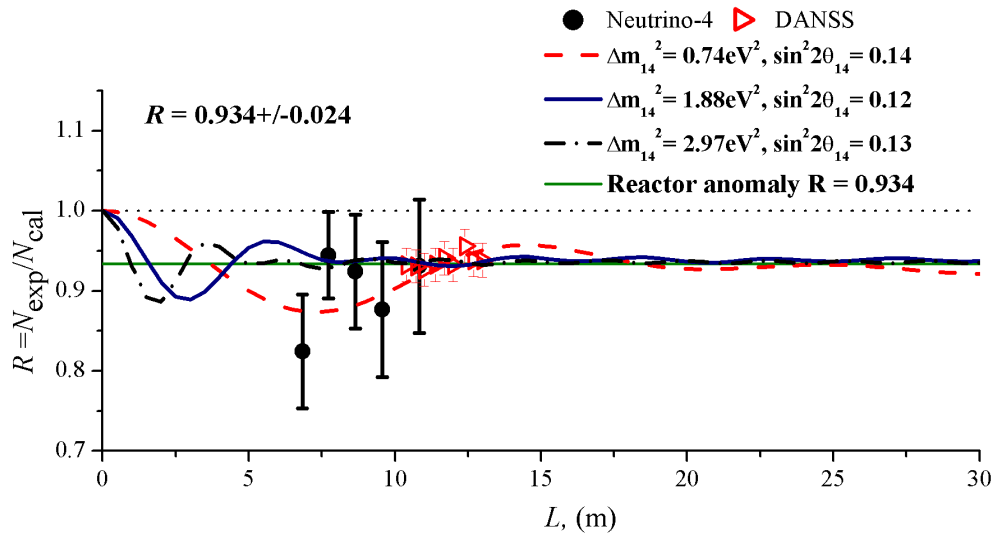

Figure 6. Fit of the sterile neutrino model parameters with experimental data of the Neutrino-4, DANSS. Curves correspond to oscillation parameters.

Further improvement of those and other experiments (PROSPECT [47], STEREO [48], SoLid [49], CHANDLER [50], CeSOX [51], IsoDAR [52], C-ADS [53], BEST [54]) seems to clarify the problem of neutrino oscillations at small range.

## 6. Hypothesis about baryonic n→n′ oscillations.

Now we return from the dark matter oscillation in neutrino sector (transition into sterile state) to dark matter oscillations in baryon sector, i.e. to n→n′ oscillation. Assume that n' mass is for instance 2÷3 MeV smaller than n. In that case transitions n→n′ would be suppressed for free neutron due to energy difference of energy states. The mass difference compensation $m_n - m_{n'}$ can be obtained in nucleus due to neutron binding energy. Then n→n′ transitions are enhanced.

At first, the general scheme of stable and unstable isotopes. (Fig. 7)

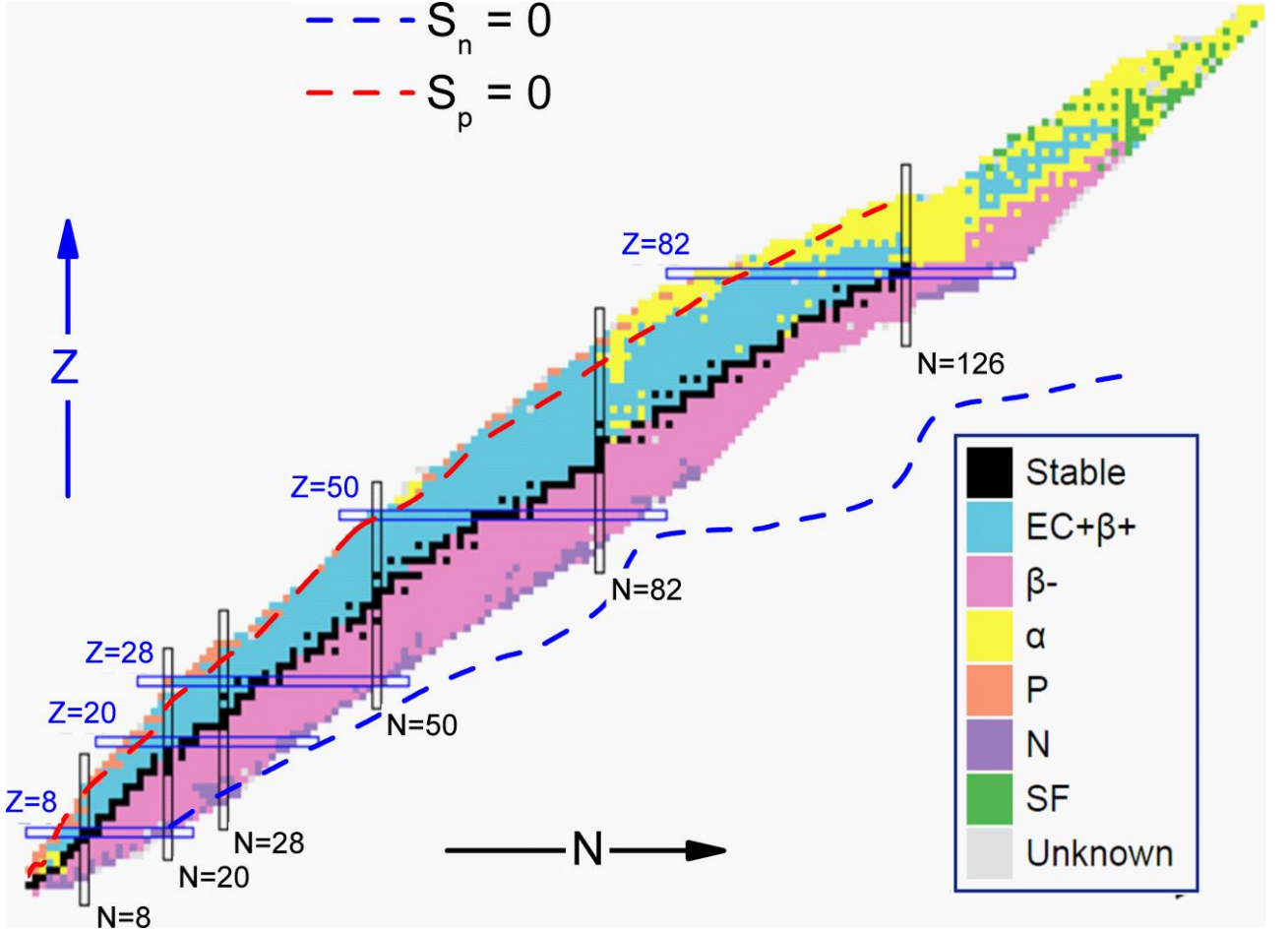

Fig.7. The central band corresponds to stable nuclei. To the left from stable nuclei are situated nuclei overloaded with protons (proton- over- filled nuclei), to the right - nuclei overloaded with neutrons, (neutron over-filled nuclei). Atomic nuclei discovered at present are shown. Line $S_p$=0 ($S_p$ - proton separation energy) is depicted by red dashed line on the left (proton drip-line) and very close to bound of observed nuclei region. $S_n$=0 ($S_n$ - neutron separation energy) is depicted by blue dashed line on the right (neutron drip-line) and it is located far from described nuclei. Notice that most neutron-emitting nuclei (lilac points in picture) has simultaneous β-decay into excited state of daughter nucleus. So, in practice, the daughter nucleus emit neutrons. Those are so called "delayed neutrons". (Adapted from [55]).

Now let us consider the neutron binding energy (neutron separation energy $S_n$) dependence on the atomic number A for stable isotopes only [56] (Fig. 8).

For majority of stable isotopes binding energy is more than 5 MeV. Exceptions are $^3$He ($S_n$ not determined and there is no corresponding point in picture) and $^9$Be ($S_n$=1.664 MeV). Considering further analysis those exceptions requires special discussion.

Neutron separation energy $S_n$ is determined as difference of binding energies of nucleus without neutron and initial nucleus [56]. Both nucleus $^3$He and $^9$Be are stable and their binding energies are positive. That stability is provided by the "last" neutron because if it is removed then daughter nuclei ($^2$He и $^8$Be) do not exist. First nucleus corresponds to two protons and second corresponds to two α-particles. Strictly speaking, here determination of energy separation is formal and obtained values should not be considered as the ones bearing physical meaning.

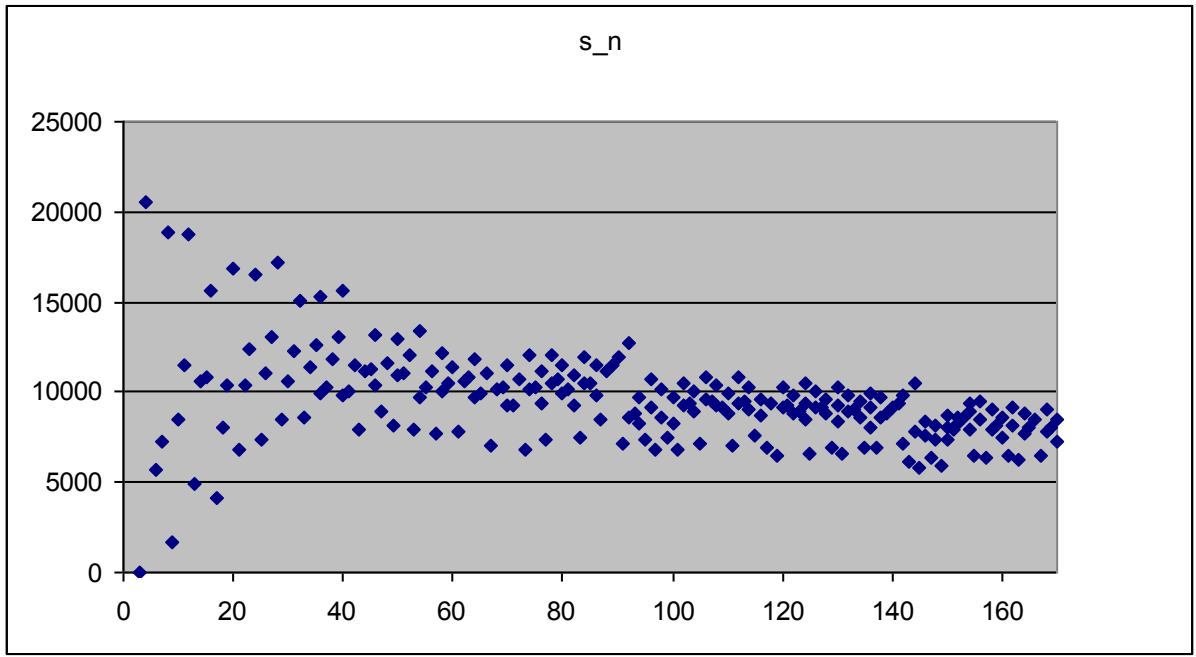

Fig.8. Neutron separation energies $S_n$ for stable nuclei with A<170.

Now let us consider the neutron binding energy (neutron separation energy $S_n$) dependence on the atomic number A for unstable isotopes. In figure 9 are shown β-decay energies $Q_β$ (blue points) and neutron separation energies $S_n$ (red points) for all β-decaying nuclei ($Q_β > 0$ with A<170).

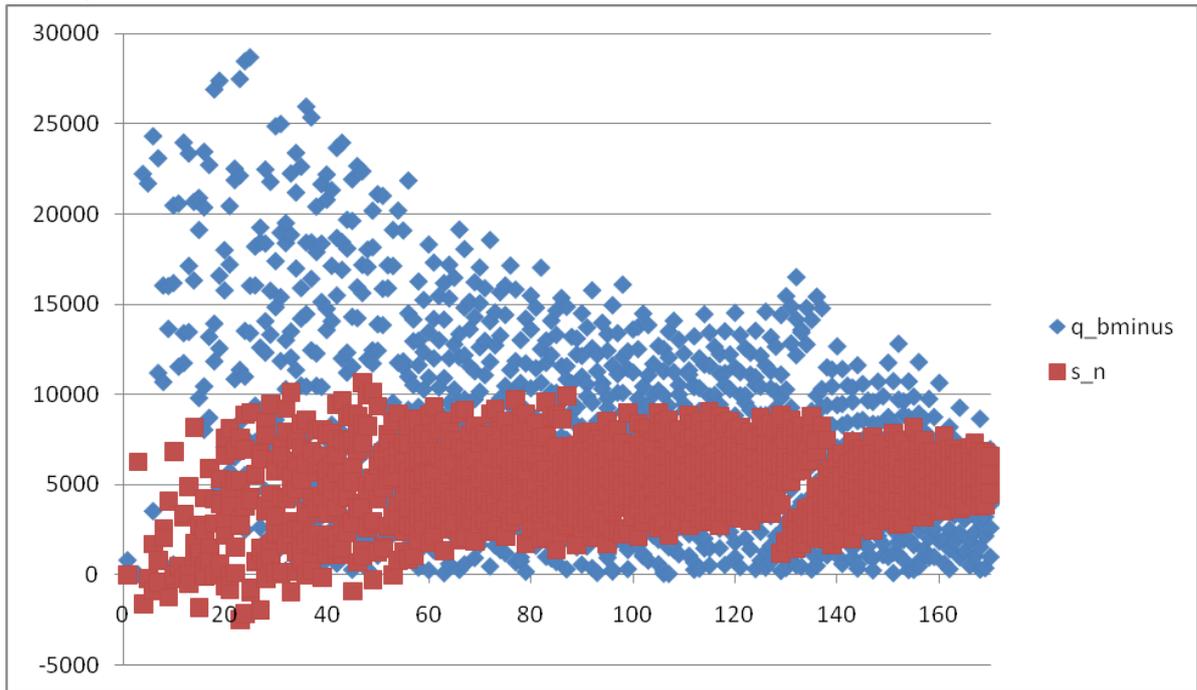

Fig.9. General scheme of β-decay energies $Q_β$ (blue points) and neutron separation energies $S_n$ (red points) in nuclei with A<170.

One can see that for A<50 there are a few nuclides which binding energy is negative, i.e. neutron nuclear decay process is possible and it competes with nuclear β-decay. That selection is shown in figure 10. However neutron decay is the strong interaction process and it has very small period about $10^{-20}$s. Hence it can not compete with β-decay which is weak interaction process.

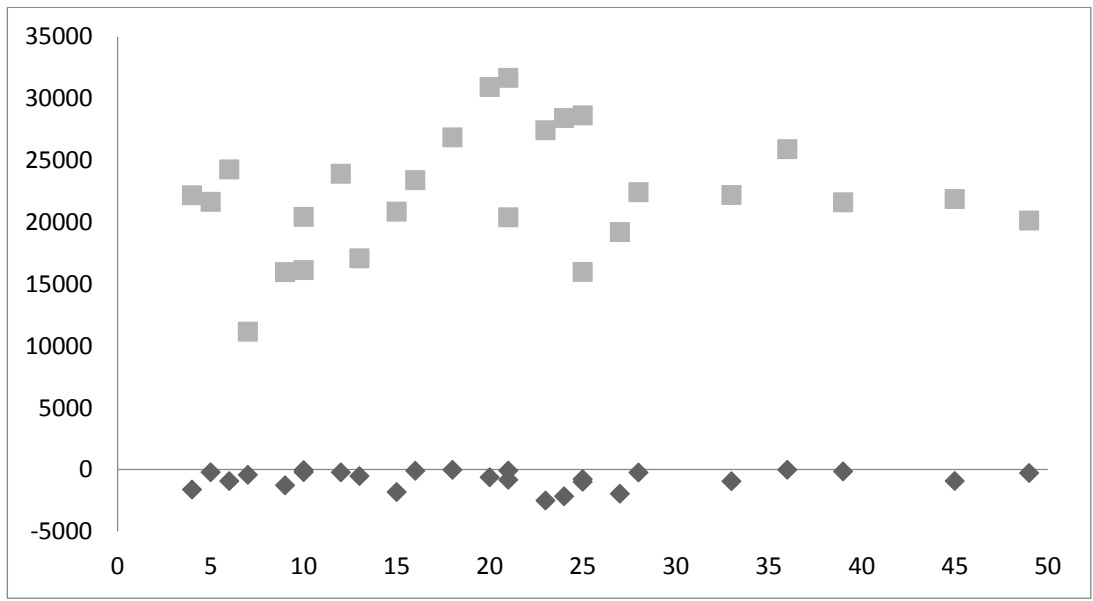

Fig.10. β-decay energy $Q$(B-) (grey squares) and neutron separation energy $S_n$ (black rhombuses) for light over-filled nuclei.

Notice that this region of A is not important for our analysis, because those isotopes do not occur in the process of $^{235}$U, $^{239}$Pu, $^{241}$Pu and $^{238}$U isotopes double fission (fig 11). Triple fission has small probability so we can remove isotopes with A<70 from our analysis.

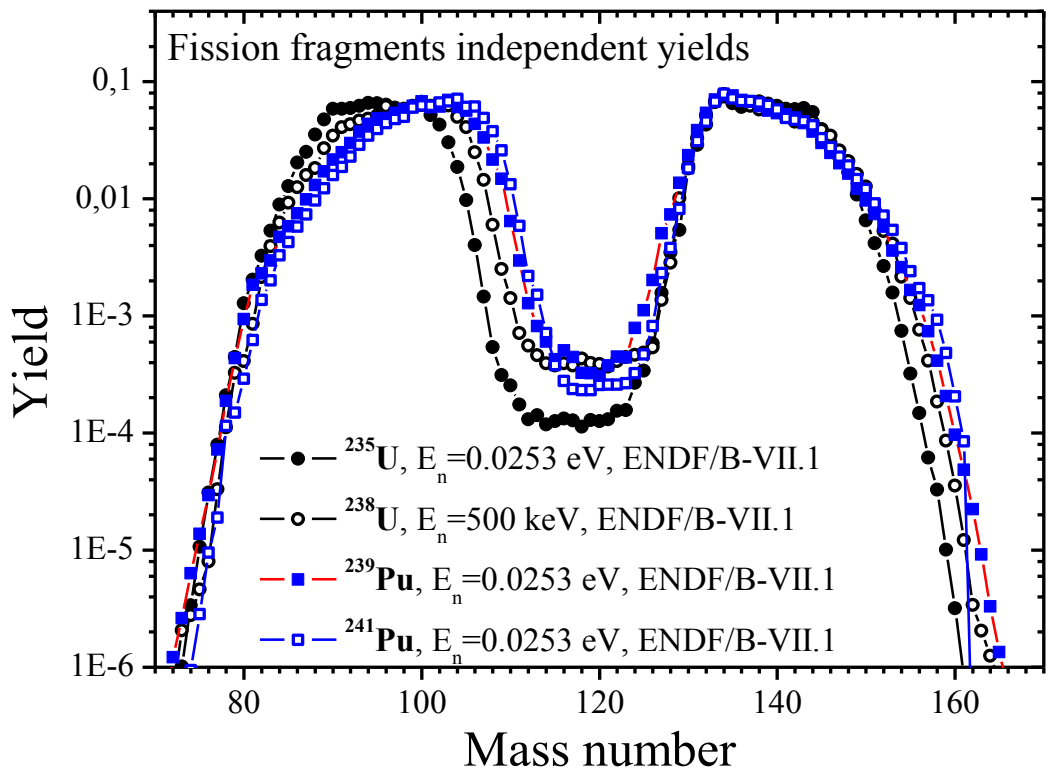

Fig. 11. Fission fragments mass diagram for isotopes $^{235}$U, $^{239}$Pu, $^{241}$Pu and $^{238}$U (Data from Evaluated Nuclear Data File ENDF/B-VII.1 [57])

Below, in figure 12 is presented interesting for us region of isotopes, which are formed in fissions. Color diagram represents fission fragments yield probability.

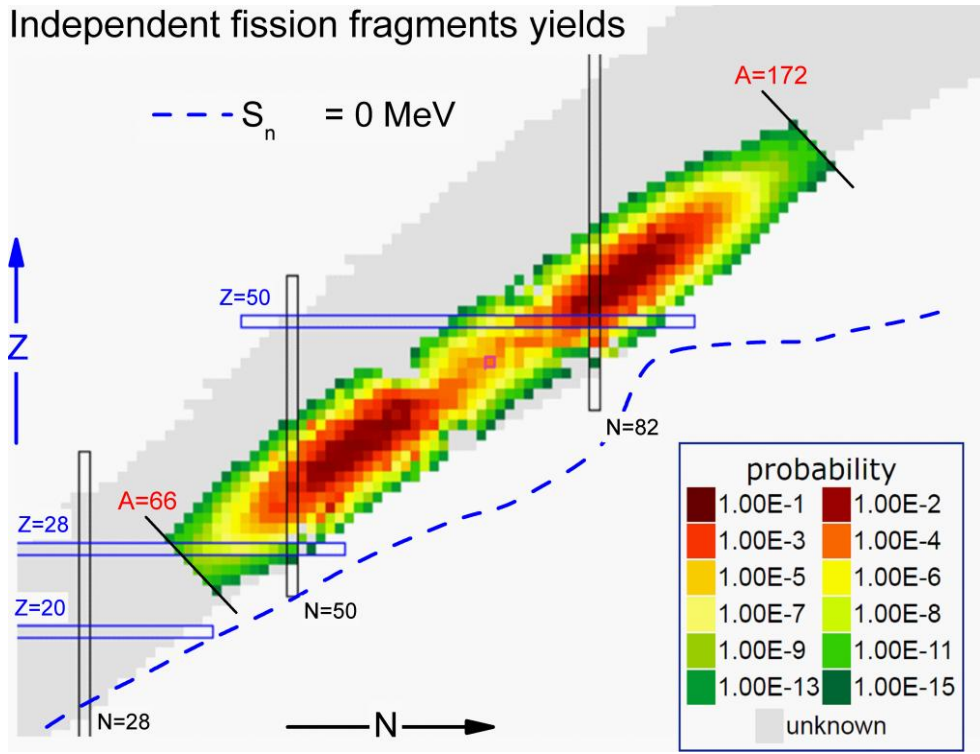

Fig. 12 Fission fragments at the nucleotid map and yeild probability. Currently observed atomic nuclei marked with grey colour. Dashed line – 0 energy of neutron separation ($S_n=0$) (Adapted from [55]).

In figure 13 is shown diagram of neutron separation $S_n$ and β-decay energies $Q_β$ for β-decaying nuclei in fission fragments region 70<A<160.

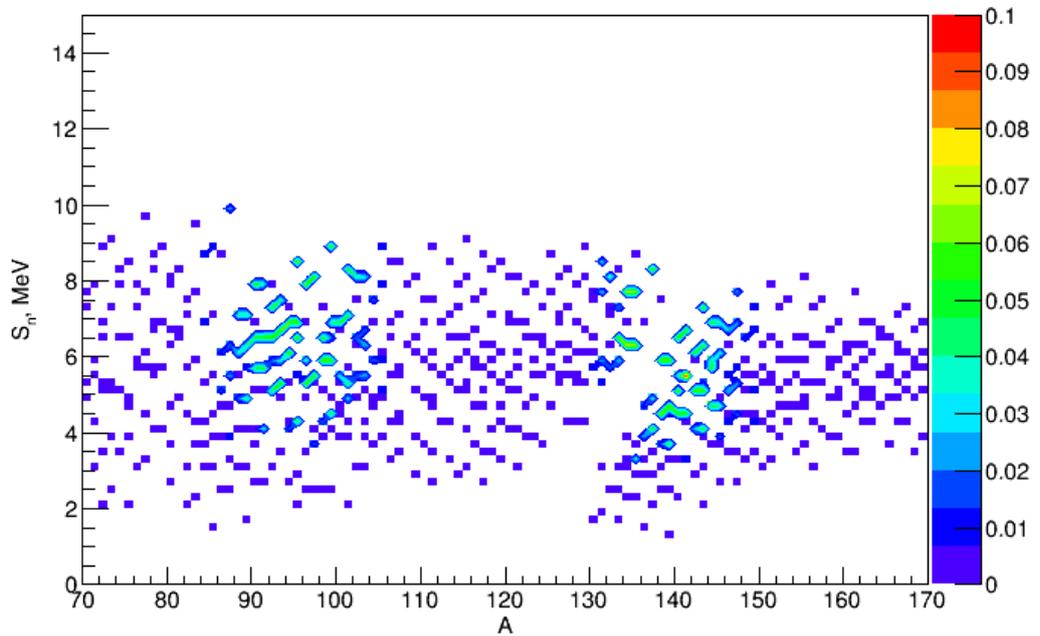

Fig.13. Neutron separation energy $S_n$ for nuclei in region of β-decaying isotopes, occurs in $^{235}$U fission. Color diagram corresponds to cumulative yield for every isotope.

One can see that there is almost no nucleotides with energies less than 2 MeV. Using the table one can find that minimal binding energy of neutron is 1.3 MeV for A=139.

The presented analysis reveals that in neutron binding energy in nuclei region with $S_n$< 2 MeV does not exist almost no β-decaying nuclei with 70< A<160. In theory, it is possible that such nuclei exist but were not observed cause of very short lifetime. In picture 14 is shown a scale of half-life period of β-active isotopes for $^{235}$U.

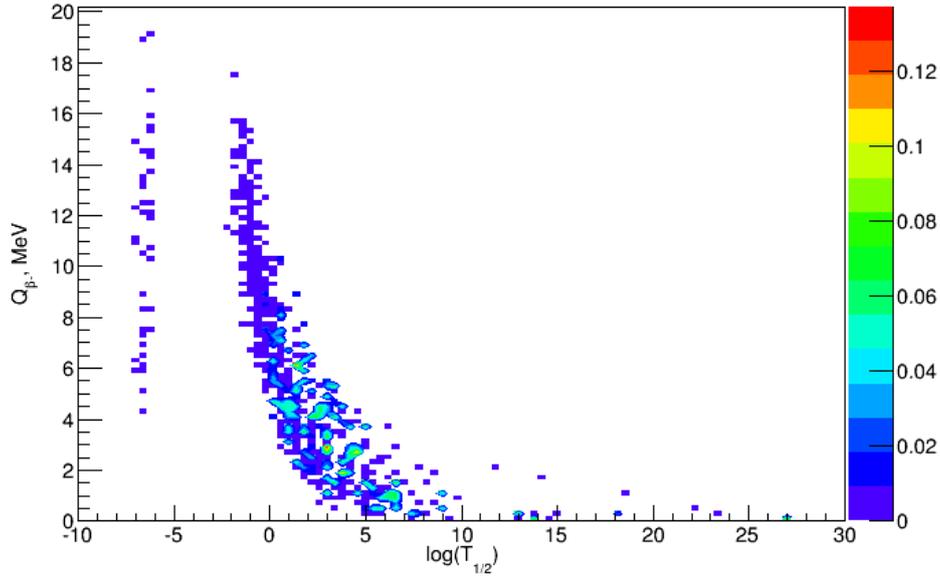

Fig.14 Energy and lifetime for nuclides, occurs in $^{235}$U fission. Color diagram corresponds to cumulative yield for each isotope.

At the bound of β-decay region and neutron-decay region is a gap of about 2-3MeV. β-decaying and neutron-decaying processes do not overlap. However that gap can be compensated if we introduce n→n′ transition.

Considering the previous analysis of mirror transitions, notice, that if we assume that mass of dark matter mirror neutron n' significantly differs from the neutron mass n, than to compensate that difference for n→n′ transition to occur the nuclear fields are required instead of magnetic, i.e. n→n′ transitions can exist in nuclei. In assumption that mirror neutron mass is less than neutron mass $m_n - m_{n'} \approx 2$ MeV, than n→n′ transitions would be suppressed for a free neutron, while their probability can be about 1% of β-decay probability to explain "neutron anomaly". In nucleus neutron is usually in a state with energy significantly smaller than energy of mirror neutron, because neutron binding energy is at level 5÷8 MeV. Transitions n→n′ in that situation are forbidden and energy conservation law ensure the stability of a nucleus. When neutron binding energy in a nucleus occurs to be equal to mass difference $m_n - m_{n'}$ (2 MeV), neutron energy levels of n and n′ coincide and n→n′ transitions become significantly enhanced. Mirror neutron leaves a nucleus.

The effect of increasing probability of n→n′ transition due to closeness of energy levels can be considered as classical resonance process and apply the Breit-Vigner function to describe it:

$$\tau_{n'}^{-1} = f(E_b) = (2\pi)^{-1} k \Gamma \left/ \left( (m_n - m_{n'} - E_e)^2 + \frac{\Gamma^2}{4} \right) \right.$$

We can normalize it so that for free neutron probability of n→n′ transition would be 1% of β-decay probability (1/880s) i.e. $1.136 \cdot 10^{-5} s^{-1}$. Then resonance function depends on neutron binding energy in a nucleus $E_b$ and determined only by mass difference $m_n - m_{n'}$. The resonance width Γ points out that lifetime in nucleus $n'$ is $10^{-21} s^{-1}$. Mirror neutron (dark matter neutron) is not held in nucleus due to no interaction with nucleons.

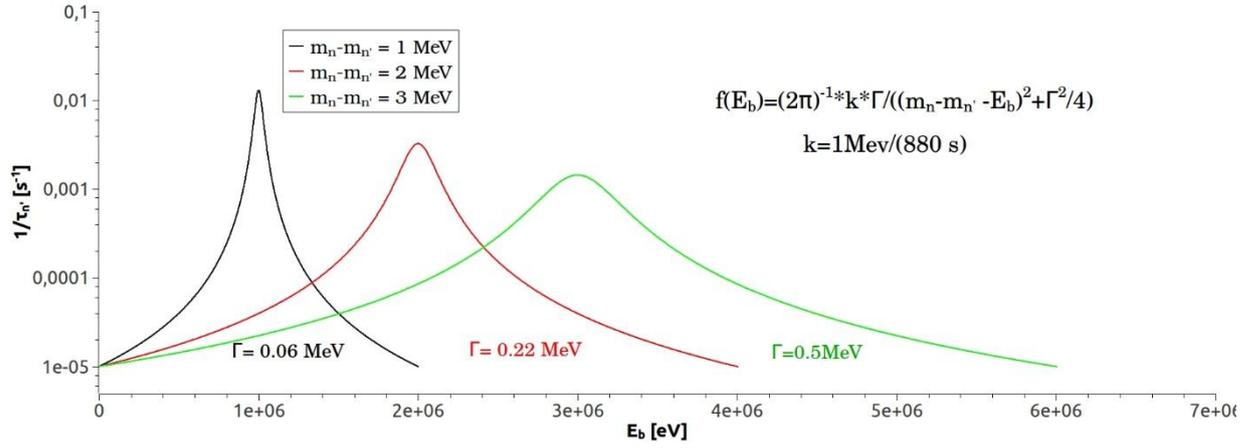

Fig.15. Effect of increasing probability of n→n′ transition due to closeness of energy levels in nucleus, $\tau_{n'}^{-1}$ dependence on binding energy of neutron in nucleus $E_b$.

Such a process can compete with β-decay, its rate will be determined by rate of $n \to n'$ oscillations, which is much less than nuclear interaction rate. It is to be noticed, that probability of $n \to n'$ oscillations is proportional to the level density of neutrons in the nucleus in $n \to n'$ resonance transition area.

Fig.16 gives a specific example for $m_n - m_{n'} = 2$ MeV. When neutron binding energy occurs to be much higher than mass difference $m_n - m_{n'}$ (Fig.16a), $n \to n'$ transition becomes energetically forbidden. When neutron binding energy is equal to mass difference $m_n - m_{n'}$ (Fig.16b), the process of $n \to n'$ transition is implemented by resonance with probability $3.1 \cdot 10^{-3}$ s$^{-1}$ instead of $1.136 \cdot 10^{-5}$ s$^{-1}$, in case of a free neutron. After realization of $n \to n'$ transition, «neutron of dark matter» escapes from the nucleus, while energy equal to mass difference $m_n - m_{n'}$ is distributed between nucleus excitation energy and kinetic energy of neutron of dark matter and recoil nucleus.

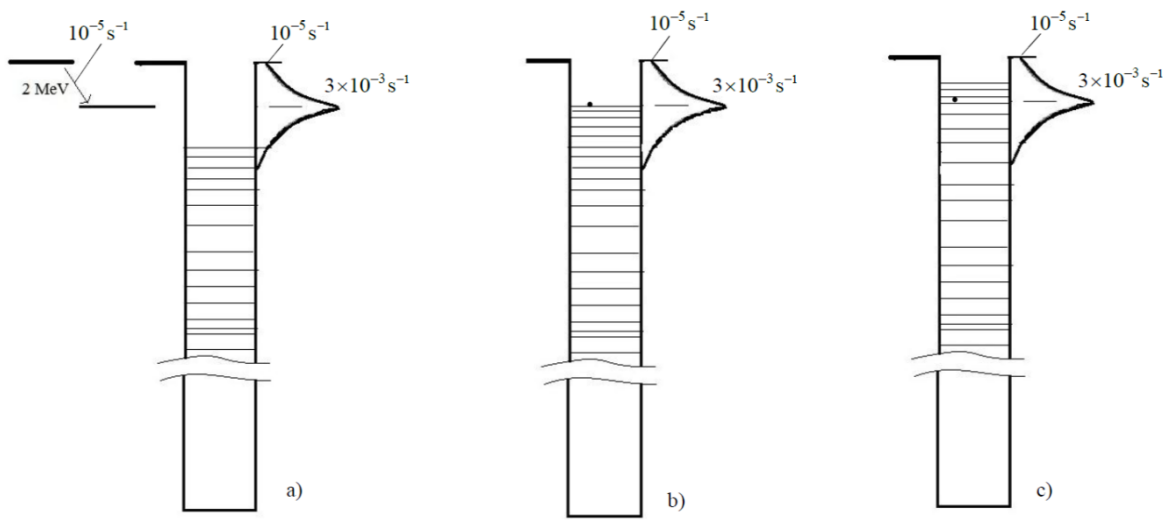

Fig . 16. n→n′ oscillations scheme: a) for a neutron with high binding energy in the nucleus ~5МэВ, providing stability for nucleus b) for a neutron in the nucleus when coupling energy compensates difference $m_n - m_{n'} \approx 2$MeV and opens up possibilities for n→n′ oscillations, c) for a neutron in the nucleus when binding energy is less than $m_n - m_{n'} \approx 2$MeV and max possibilities are opened for n→n′ oscillations.

Principally the process can be experimentally discovered according to energy of nucleus excitation emission and one can calculate dark matter neutron mass. When neutron binding energy is less than mass difference (Fig.16c), the process of $n \rightarrow n'$ transition should be calculated taking into account neutron levels density in the region of resonance amplification.

Now we go back to conclusions of proposed scheme for "reactor antineutrino anomaly". Pushing the bound of neutron radioactivity process (decay through mirror neutron channel) toward the β-radioactivity bound we obtain the process which competes with β-decay. If it is not taken into consideration in calculations then the calculated β-particles yield and hence antineutrino yield are overestimated, i.e. we obtain the effect of "reactor antineutrino anomaly". If we assume that in binding energy region from 2 to 4 MeV occurs the decreased yield of β-particles, then there could be a possibility to quantitatively explain the "reactor antineutrino anomaly". Notice that correction of antineutrino yield for low binding energy isotopes mostly affects upper part of antineutrino spectrum, i.e. in the region detectable by inverse beta decay reaction. Antineutrino deficiency is 6.6±2.4% and neutron anomaly is 1.0 ± 0.3%.

On the whole, one can conclude that detailed calculations of the proposed model will require one free parameter: difference of masses $m_n - m_{n'}$, if probability of n→n′ oscillations for a free neutron normalize on «neutron anomaly», 1%. Having attained explanation for 6.6% of neutrino anomaly, one can determine difference of masses $m_n - m_{n'}$ and thus neutron mass. Such is our naive scheme.

To implement it, there is no need to make detailed calculations of the spectrum of the reactor antineutrino. It is sufficient to make comparable calculations of antineutrino yield: without accounting for n→n′ oscillations effect in the nucleus, i.e. as usual, and taking account of effect of n→n′ oscillations in the nucleus. The integral of $\beta$-decayers yield gives the number of antineutrino per fission event. To calculate the number of antineutrino registered with a detector, based on the reaction of reverse $\beta$-decay, one should take into account the threshold of this reaction 1.8 MeV. Therefore, one should exclude isotopes with $Q_\beta$<1.8 MeV from the list. Thus, we obtain the number of antineutrino registered by the detector per fission event, for example, for $^{235}$U. The same procedure can be repeated, considering that antineutrino yields may be made somewhat less as a result of n→n′ transition. This factor is $\tau_\beta^{-1}/(\tau_\beta^{-1}+\tau_{n'}^{-1})$ i.e. ratio of probability of $\beta$-decay to the total decay probability, where $\tau_{n'}^{-1}$ is probability of n→n′ transition. In the second calculation we will obtain antineutrino yield per fission even, decreased at the expense of n→n′ transition. It should be noticed that in tables presented experimental data $\tau_{exp}^{-1}$, which considered to be β-decay probability. But if we assume that competing process $\tau_{n'}^{-1}$ exists, than sum of probabilities is actually measured in experiment, i.e. $\tau_{exp}^{-1}=\tau_\beta^{-1}+\tau_{n'}^{-1}$. In that scheme correction factor is:

$$\tau_\beta^{-1}/(\tau_\beta^{-1}+\tau_{n'}^{-1})=(\tau_{exp}^{-1}-\tau_{n'}^{-1})/\tau_{exp}^{-1}$$

Table 1 presents yields of the number of antineutrino ($N_{\bar{\nu}}$ / *fission*) per event of different fission isotopes $^{235}$U, $^{239}$Pu, $^{241}$Pu, $^{238}$U, yields of antineutrino with energy higher than 1.8 MeV, i.e. higher than registration threshold of the detector based on reaction of reverse $\beta$-decay.

Table 1. Yields of the number of antineutrino ($N_{\bar{\nu}}$ / fission) through fission event for different isotopes

| Isotope  Threshold | $^{235}$U  $N_{\bar{\nu}}$ / fission | $^{239}$Pu  $N_{\bar{\nu}}$ / fission | $^{241}$Pu  $N_{\bar{\nu}}$ / fission | $^{238}$U  $N_{\bar{\nu}}$ / fission |
|---|---|---|---|---|
| 0 MeV | 5.84 ± 0.11 | 5.21 ± 0.11 | 5.95 ± 0.11 | 7.20 ± 0.14 |
| 1.8 MeV | 4.05 ± 0.07 | 3.44 ± 0.07 | 4.16 ± 0.07 | 5.39 ± 0.11 |

Table 2 presents results of calculations on the effect of «reactor neutrino anomaly» $\Delta N_\nu / N_\nu$ (%) for different dividing isotopes $^{235}$U, $^{239}$Pu, $^{241}$Pu, $^{238}$U and for mass difference $m_n - m_{n'} = 1 \div 3.75$ MeV.

Table 2. Results of calculations on effect of «reactor neutrino anomaly» $\Delta N_\nu / N_\nu$ (%)

| Isotope  ($m_n - m_{n'}$) | $^{235}$U  $\Delta N_{\bar{\nu}} / N_{\bar{\nu}}$ (%) | $^{239}$Pu  $\Delta N_{\bar{\nu}} / N_{\bar{\nu}}$ (%) | $^{241}$Pu  $\Delta N_{\bar{\nu}} / N_{\bar{\nu}}$ (%) | $^{238}$U  $\Delta N_{\bar{\nu}} / N_{\bar{\nu}}$ (%) |
|---|---|---|---|---|
| 1 MeV | 0.56 ± 0.08 | 0.53 ± 0.04 | 0.41 ± 0.02 | 0.43 ± 0.06 |
| 2.5 MeV | 2.70 ± 0.10 | 2.64 ± 0.08 | 2.05 ± 1.72 | 2.11 ± 0.10 |
| 3 MeV | 3.75 ± 0.16 | 3.55 ± 0.14 | 2.84 ± 0.10 | 2.86 ± 0.14 |
| **3.25 MeV** | **5.39 ± 0.20** | **5.12 ± 0.17** | **4.12 ± 0.12** | **4.12 ± 0.18** |
| 3.75 MeV | 9.36 ± 0.29 | 10.01 ± 0.27 | 8.41 ± 0.20 | 7.21 ± 0.27 |

From table 2 one can conclude that «antineutrino deficit» at 5% level is attained at $m_n - m_{n'} \approx 3.25$ MeV. One should point out that if 'this scheme is true, in calculation of efficiency of antineutrino detector for cross-section of the reaction of reverse $\beta$-decay, one needs to use beam lifetime of neutron $\tau_n = 887.7 \pm 3.1 c$, rather than $\tau_n$ from experiments with UCN trap. This will diminish "antineutrino deficit" from c 6.6% to 5.6%. Thus, we prefer to choose the estimation $m_n - m_{n'} \approx 3.25$ MeV.

One should emphasize, that in decay of a free neutron along the channel $n \to n'$ energy must be emitted equal to difference of masses $m_n - m_{n'}$, the experiment [38] should be repeated by extending energy range of measuring

Finally, direct experimental test of dark matter channel n→n′ have to be carried out. Considering dark matter neutron to be unobservable, we can rely only on observing so called "mass leak" at process of β-decay of cumulative isotopes. The process of formation "antineutrino deficiency" have to go along with "mass leaking into dark matter channel". Numeric calculations can be easily carried out using developed calculation scheme. In table 3 dark matter neutron yields are listed for various decaying isotopes. The estimation reveals that dark matter neutron yield is 0.25 u for fission process. Experimental observation mass leak of 0.25u at the background of 235u is impossible, especially considering the fact that the process occurs along with neutron leak, capturing of neutrons and neutrino emission. Nevertheless, majority of process can be taken into account in calculation. For now, it is unclear if the method of relative weigh of new and used nuclear fuel element can be used.

Table 3 Dark matter yield per fission

| Isotope $(m_n - m_{n'})$ | $^{235}$U $N_{n'}\cdot(fission)^{-1}$ | $^{239}$Pu $N_{n'}\cdot(fission)^{-1}$ | $^{241}$Pu $N_{n'}\cdot(fission)^{-1}$ | $^{238}$U $N_{n'}\cdot(fission)^{-1}$ |
|---|---|---|---|---|
| 1 MeV | 0.024 ± 0.004 | 0.019 ± 0.002 | 0.018 ± 0.001 | 0.024 ± 0.004 |
| 2.5 MeV | 0.111 ± 0.006 | 0.092 ± 0.004 | 0.086 ± 0.003 | 0.115 ± 0.007 |
| 3 MeV | 0.152 ± 0.010 | 0.122 ± 0.007 | 0.118 ± 0.006 | 0.154 ± 0.011 |
| **3.25 MeV** | **0.218 ± 0.013** | **0.176 ± 0.009** | **0.171 ± 0.007** | **0.222 ± 0.014** |
| 3.75 MeV | 0.374 ± 0.023 | 0.341 ± 0.016 | 0.348 ± 0.013 | 0.384 ± 0.025 |

Thus dark matter neutrons yield per fission is estimated to be 0.2 u.

Another variant is to consider schemes of n→n′ channel decay and determine final isotopes, which do not appear in usual scheme of β-decay of cumulative isotopes, but correlates with them.

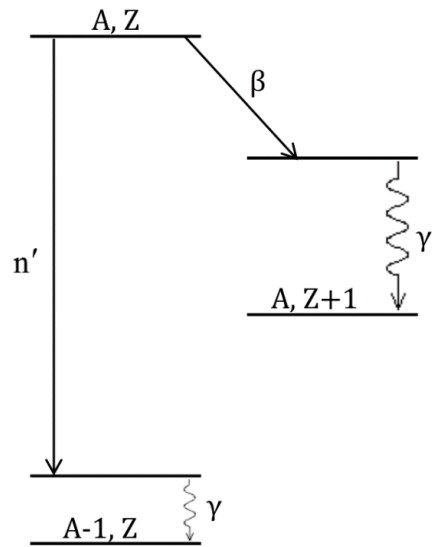

Рис 17. Scheme of discussed $\beta$-decay and n′-decay.

In figure 17 is shown discussed scheme of competing β-decay and $n'$-decay. We can not observe $n'$ dark matter neutron but we can try to search for A-1,Z in final state n′-decay, which can occur in correlation with isotope A,Z+1 from $\beta$-decay. For now, we can suggest only β-decaying isotopes, where process of dark matter neutron yield have most probability (Fig. 18). Isotopes, which have relatively high probability of emission of dark matter neutron, are listed in Table 4.

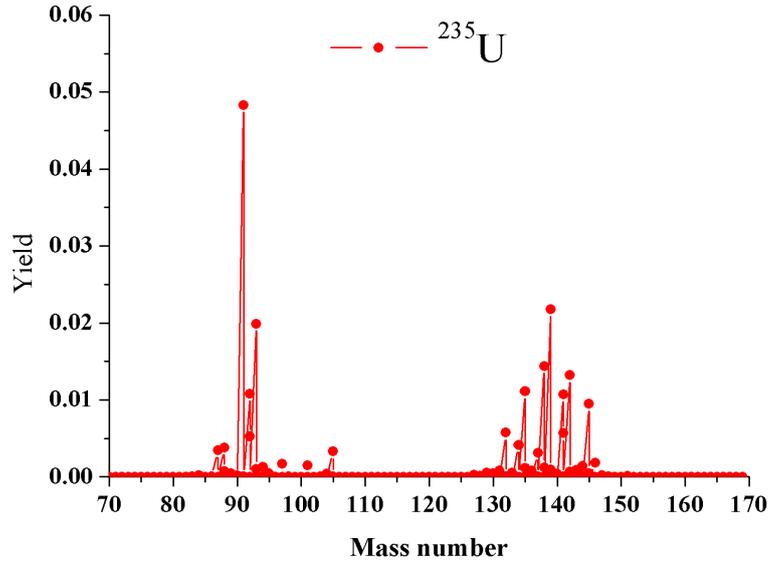

Fig. 18. Yield of cumulative isotopes in fissions of $^{235}$U nuclei, in which dark matter neutrons can occur.

Table 4.

| Parameters / Isotope | $T_{1/2}$ of isotope (A, Z) | Cumulative yield (A, Z), $10^{-2}$ | Relative yield $n'$, % | $T_{1/2}$ of isotope (A – 1, Z) from $n'$-decay | Cumulative yield (A-1, Z), $10^{-2}$ | $T_{1/2}$ of isotope (A, Z+1) from β-decay | Cumulative yield (A, Z+1), $10^{-2}$ |
|---|---|---|---|---|---|---|---|
| $^{91}$Sr | 9.65 h | 5.83 | 83 | $^{90}$Sr 28.9 y | 5.78 | $^{91}$Y 58.51 d | 5.83 |
| $^{139}$Ba | 1.38 h | 6.41 | 34 | $^{138}$Ba Stable | 6.77 | $^{139}$La Stable | 6.41 |
| $^{93}$Y | 10.18 h | 6.35 | 31 | $^{92}$Y 3.54 h | 2.21 | $^{93}$Zr 1.53·10$^6$ y | 6.35 |
| $^{138}$Cs | 33.41 min | 6.71 | 21 | $^{137}$Cs 30.17 y | 6.19 | $^{138}$Ba Stable | 6.77 |
| $^{142}$La | 1.52 h | 5.85 | 23 | $^{141}$La 3.92 h | 5.85 | $^{142}$Ce Stable | 5.85 |
| $^{135}$I | 6.57 h | 6.28 | 18 | $^{134}$I 52 min | 7.83 | $^{135}$Xe 9.14 h | 6.54 |
| $^{92}$Y | 3.54 h | 6.01 | 18 | $^{91}$Y 58.51 d | 5.83 | $^{92}$Zr Stable | 6.02 |

Cumulative yield of daughter isotopes for two possible channels $n'$-decay and β-decay listed in corresponding columns. Expected yield ratio is shown in column "relative yield $n'$". Analysis of listed cumulative yields reveals that experimental data do not satisfy that ratio. Hence, experimental confirmation of discussed ideas of possible existence of mirror dark matter neutron with mass difference $m_n - m_{n'} \approx 3$ MeV was not obtained.

**Conclusion**

It is conventional to start a serious discussion about an effect or experimental data discrepancy than it reaches 5σ, e.g. resonances in high energy physics. However, usually the discussion on ideas starts from 3σ level considering it enough to start an analysis of the interpretation, yet it is not source of final conclusions. Creativity should not and can not be stopped.

In this work we discussed two so-called anomaly: neutron and reactor. Each of them is at confidence level of 3σ. ("antineutrino deficiency" is 6.6±2.4% and "neutron anomaly" is 1.0±0.3%). Clearly, there is high possibility that after obtaining new results and performing detailed calculation those problems would be solved without employing any new physics. That is the most common scenario. However, such work has to be finished.

The peculiarity of the proposal in this article was that both anomalies can be explained by single phenomenon of oscillation in the baryon sector between neutron and dark matter neutron with mass somewhat less than an ordinary neutron mass. At current level we could not find any confirmation of proposed scheme and existence of mirror dark matter neutron with mass difference $m_n - m_{n'} \approx 3$ MeV, using data of fission fragments yields. The result of the analysis is the conclusion that for mirror neutrons the region of the mass difference $m_n - m_{n'} \geq 3$ MeV is closed. The region of the mass difference $m_n - m_{n'} \leq 2$ MeV turned out to be not closed, because there are practically no nuclides with neutron binding energies below 2 MeV.

However, the possibility of observing dark matter in laboratory experiment is rather intriguing.

**Acknowledgments**
This work was supported by the Russian Science Foundation (Project № 14-22-00105).